\documentclass{PoS}

\PoS{PoS(LAT2005)120}

\title{Locality and Scaling of Quenched Overlap Fermions 
       \thanks{This research is supported in part by the U.S. Department of
               Energy under grants DE-FG05-84ER40154, DE-FG02-02ER45967, and
               DE-FG02-95ER40907.}  
      }

\ShortTitle{Locality and Scaling of Quenched Overlap Fermions}

\author{$\chi$QCD Collaboration: 
        \speaker{Terrence Draper}$^{a}$,
        Nilmani Mathur$^{a,b}$, Jianbo Zhang$^{c}$,
        Andrei Alexandru$^{a}$, Ying Chen$^{d}$, Shao-Jing Dong$^{a}$, 
        Ivan Horv\'{a}th$^{a}$, Frank Lee$^{e}$, and Sonali Tamhankar$^{a,f}$\\ \\
        {$^{a}$} Department of Physics and Astronomy, 
                 University of Kentucky, 
                 Lexington, KY 40506, USA \\
        {$^{b}$} Jefferson Lab, 
                 12000 Jefferson Avenue, 
                 Newport News, VA 23606, USA \\
        {$^{c}$} CSSM and Department of Physics,
                 University of Adelaide, 
                 Adelaide, SA 5005, Australia \\
        {$^{d}$} Institute of High Energy Physics, 
                 Academia Sinica, 
                 Beijing 100039, P.R. China \\
        {$^{e}$} Center for Nuclear Studies, 
                 Dept.\ of Physics, 
                 George Washington Univ.,
                 Washington, DC 20052, USA \\
        {$^{f}$} Department of Physics, Hamline University, St. Paul, MN 55104, USA \\
        E-mail: \email{draper@pa.uky.edu}}

\abstract{The overlap fermion offers the tremendous advantage of exact chiral
          symmetry on the lattice, but is numerically intensive.  This can be
          made affordable while still providing large lattice volumes, by using
          coarse lattice spacing, given that good scaling and localization
          properties are established. Here, using overlap fermions on quenched
          Iwasaki gauge configurations, we demonstrate directly that the
          overlap Dirac operator's range is comfortably small in lattice units
          for each of the lattice spacings 0.20 fm, 0.17 fm, and 0.13 fm (and
          scales to zero in physical units in the continuum limit).  In
          particular, our direct results contradict recent speculation that an
          inverse lattice spacing of $1\,{\rm GeV}$ is too low to have
          satisfactory localization.  Furthermore, hadronic masses (available
          on the two coarser lattices) scale very well.}

\FullConference{XXIIIrd International Symposium on Lattice Field Theory\\
                25-30 July 2005\\
                Trinity College, Dublin, Ireland}

\begin{document}

\section{Locality}

In the last several years the use of the overlap fermion has become more
popular because the conceptual and technical clarity that results from its
exact chiral symmetry on the lattice is seen to overcome its superficially
higher computational cost as compared to conventional formulations of the
fermion.  Furthermore, the disparity in numerical intensity can be mitigated by
using coarse lattice spacing, given that good scaling and localization
properties are established.

Hern\'{a}ndez, Jansen, and L\"{u}scher~\cite{Her99} showed numerically that
Neuberger's overlap operator is local (using the Wilson gauge action on fine
lattices).  Recently, Golterman, Shamir, and Svetitsky~\cite{Gol05} have
speculated that overlap simulations with a cutoff of $1\,{\rm GeV}$ (such
as~\cite{Che04}) might have a range as long as 4 lattice units, and thus be
afflicted by unphysical degrees of freedom as light as $0.25\,{\rm GeV}$.  Here
we show directly that such is not the case; the range is about 1 lattice unit
(in Euclidean distance or 2 units of ``taxi-driver'' distance).  All is well.

\subsection{Lattice Details}

We use the renormalization-group-improved Iwasaki~\cite{Iwa85} gauge action, on
three different lattices; for each, the lattice size, lattice spacing, and
number of configurations used are tabulated in Table~\ref{Table:lattices}.

\begin{table}[hb] 
  \begin{center}
    \begin{tabular}{llr}
      $N_s \times N_t$   & $a ({\rm fm})$ &  $N_{\rm cfg}$ \\
      \hline
      $16^3\times 28$    & $0.20$         & $300/10$ \\
      $20^3\times 32$    & $0.17$         & $ 98/10$ \\
      $28^3\times 44$    & $0.13$         & $   /10$ \\
      \hline
    \end{tabular}
    \caption{\label{Table:lattices} Lattice size, lattice spacing, number of
             configurations (for scaling/locality). }
  \end{center}
\end{table}

For the associated scaling study of hadron masses, we use the overlap
fermion~\cite{Neu98} and massive overlap operator~\cite{Ale00}
\begin{eqnarray*}
        D(m_0) 
        & = &
        (\rho + \frac{m_0a}{2}) + (\rho - \frac{m_0a}{2} ) \gamma_5 \epsilon (H)
\end{eqnarray*}
where $\epsilon (H) = H /\sqrt{H^2}$, $H = \gamma_5 D_w$, and $D_w$ is the
usual Wilson fermion operator, except with a negative mass parameter $-\rho =
1/2\kappa -4$ in which $\kappa_c < \kappa < 0.25$; we take $\kappa = 0.19$ in
our calculation which corresponds to $\rho = 1.368$.  For the locality study,
we set $m_0=0$ to look at the properties of the massless operator $D(0)$.
Complete details are described in~\cite{Che04}.

\subsection{Locality as Measured by Taxi-Driver Distance}

It is convenient for formal reasons to discuss locality in terms of
``taxi-driver'' distance~\cite{Her99}. 
\begin{equation}
  r_{\rm TD} = || x-y||_1 = \sum_{\mu} |x_\mu - y_\mu|
\end{equation}
The locality of the overlap operator is then studied by plotting the quantity
$|D(r)|$ ($f(r)$ in the notation of~\cite{Her99}) as a function of the
taxi-driver distance for a localized source,
$\psi_{\alpha}(x)=\delta(x)\delta_{\alpha\beta}$ for fixed Dirac-color index
$\beta$.
\begin{equation}
  |D(r)| = \max \{ ||D\psi(x)|| \,\, | \,\, \sum_{\mu} x_{\mu}=r \}
\end{equation}
For large $r$, the kernel of the Dirac-overlap operator decays exponentially
with decay rate $\nu=r_{\rm ov}^{-1}$, where $r_{\rm ov}$ is the range
(characteristic decay distance) measured in lattice units.

\subsection{Results}

\begin{figure}[ht]
  \vspace{0cm}
  \begin{center}
  \includegraphics[angle=0,width=0.5\hsize]{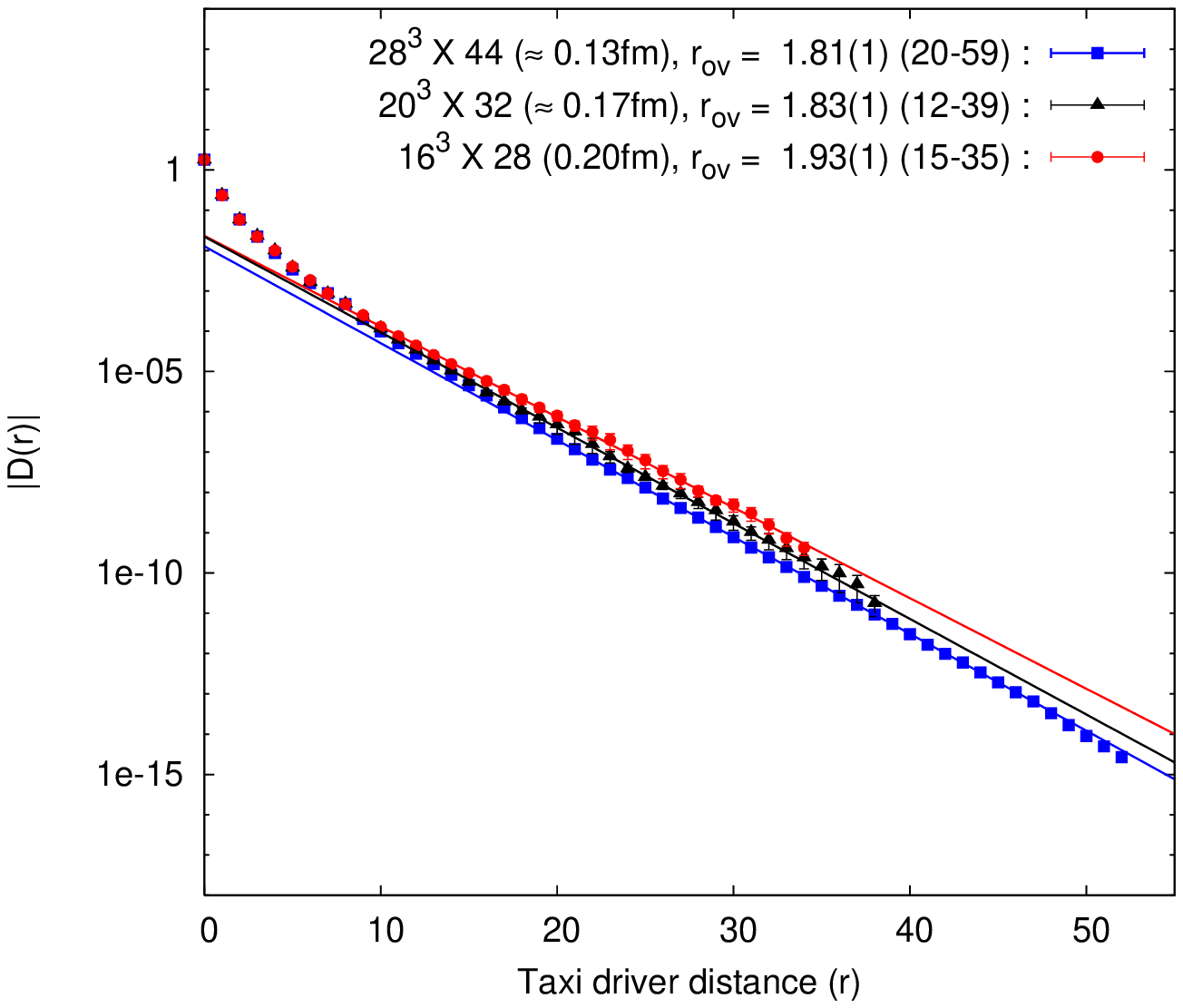}%
  \includegraphics[angle=0,width=0.5\hsize]{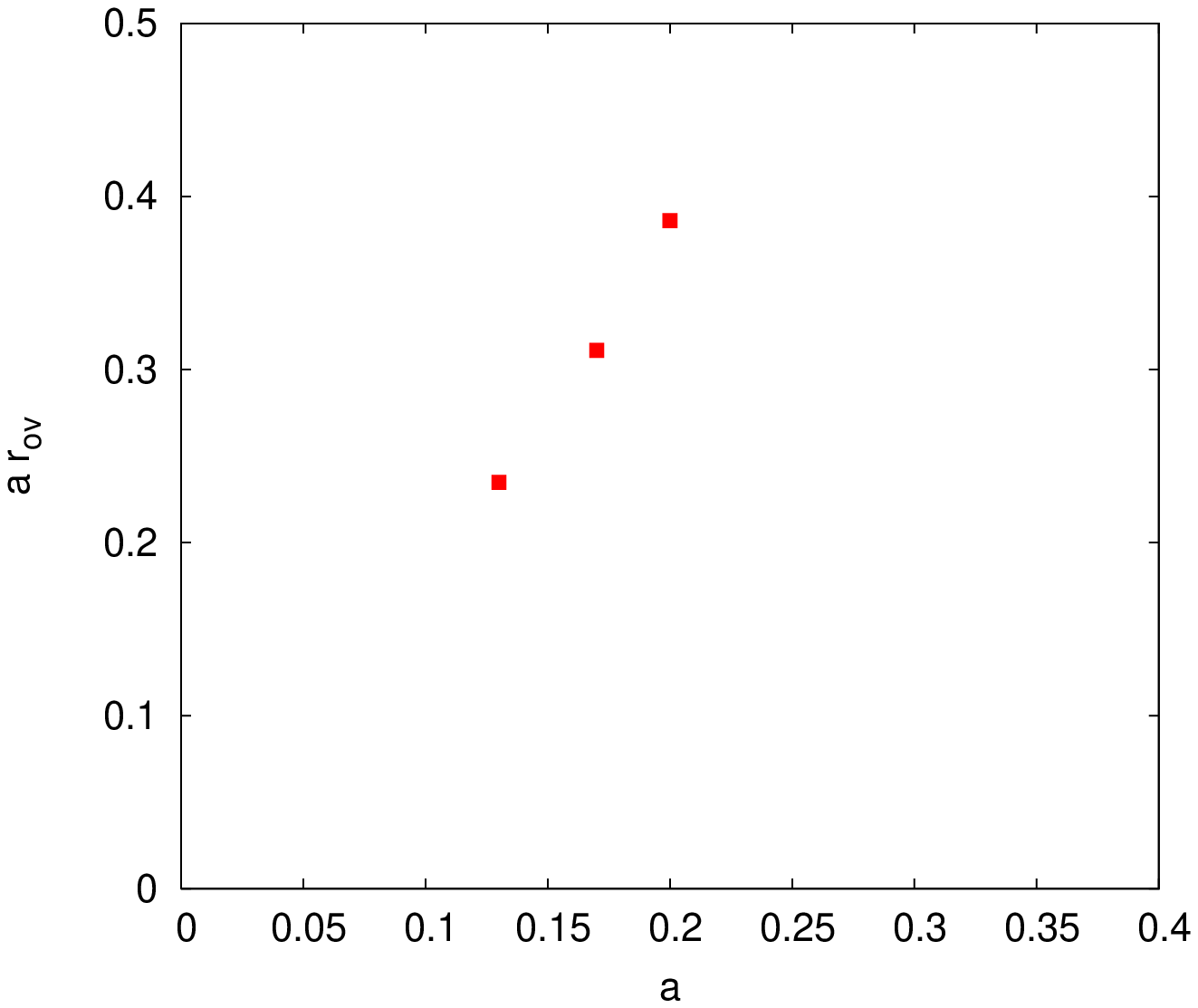}
  \vspace{-0.5cm}
  \caption{\label{Fig:locality} Left: For each of three lattice spacings, the
           expectation value of $|D(r)|$ as a function of the taxi-driver
           distance, $r$.  For large $r$, $|D(r))|$ falls exponentially, with
           range $r_{\rm ov}$.  Fitted values of $r_{\rm ov}$ are shown with
           fit intervals.  Right: Taxi driver range in physical units (fm) as a
           function of lattice spacing.  The range is small even at coarse
           lattice spacing and trends to zero in the continuum limit.  }
  \vspace{0cm}
  \end{center}
\end{figure}

In the left pane of Fig.~\ref{Fig:locality}, we plot $|D(r)|$ as a function of
taxi-driver distance for each of three lattice spacings.  At large distances,
we fit to an exponentially decreasing function to extract the range $r_{\rm
ov}$.  These are tabulated in Table~\ref{Table:locality}.  We note that our
results are consistent with the results of Hern\'{a}ndez, Jansen and
L\"{u}scher~\cite{Her99} on finer lattices for the overlap operator
($\rho=1.4$) with Wilson action at $\beta=6.0$, $6.2$, and $6.4$ where they
find $\nu=r_{\rm ov}^{-1}=0.49$.

\begin{table}[ht]
  \begin{center}
    \begin{tabular}{ll}
      $a$   &  $r_{\rm ov}$ \\
      \hline
      $0.20\,{\rm fm}$ & $1.93(1)$ \\
      $0.17\,{\rm fm}$ & $1.83(1)$ \\
      $0.13\,{\rm fm}$ & $1.81(1)$ \\
      \hline
    \end{tabular}
    \caption{\label{Table:locality} The range (taxi driver metric) for three
             lattice spacings.  It is less than two lattice units on a lattice
             as coarse as $0.20\,{\rm fm}$.}

  \end{center}
\end{table}

Furthermore, it is gratifying to see that even for our coarsest lattice,
$0.20\,{\rm fm}$, the {\em measured\/} range is less than two lattice units.
In the right pane of Fig.~\ref{Fig:locality} we plot the range in physical
units as a function of lattice spacing; it trends to zero in the continuum
limit.

\subsection{Locality as Measured by Euclidean Distance}

We conclude that it is perfectly acceptable to simulate overlap fermions with
lattice spacing as coarse as $0.20\,{\rm fm}$, since for this we find that the
range is not greater than two lattice units when measured in taxi-driver
distance.  In fact, the situation is even better than it seems.  To see this,
we consider the more familiar standard Euclidean metric
\begin{equation}
  r_{\rm E} = || x - y ||_2 = \sqrt{\sum_{\mu} |x_\mu - y_\mu|^2}
\end{equation}

\begin{figure}[ht]
  \vspace{0cm}
  \begin{center}
  \includegraphics[angle=0,width=0.6\hsize]{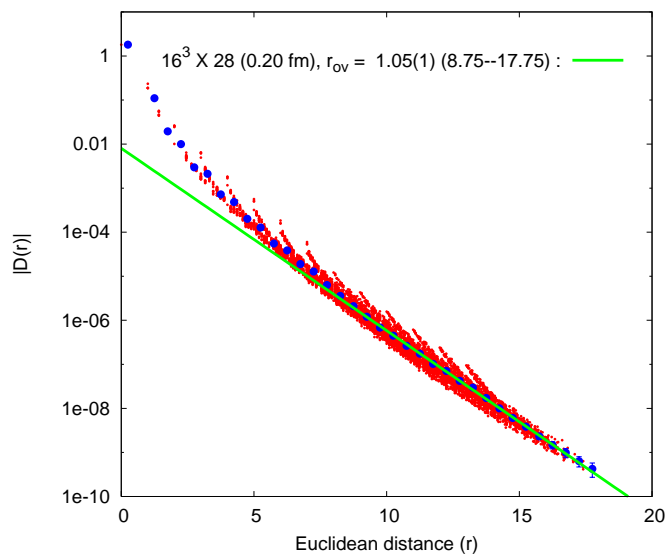}
  \vspace{0cm}
  \caption{\label{Fig:Euclidean} For the $16^3\times 28$ lattice, the
           expectation value of $|D(r)|$ as a function of Euclidean distance
           $r$ (for data cut to remove wrap-around effects).  Data is plotted
           at each Euclidean distance in red.  The blue points are averages
           within bins of width $\Delta{r}=0.5$.  The exponential tail is then
           fit over the longest interval possible with acceptable $\chi^2$, as
           shown by the green line which has inverse slope $r_{\rm
           ov}=1.05(1)$, the range.}
  \vspace{0cm}
  \end{center}
\end{figure}

In Fig.~\ref{Fig:Euclidean} we plot $D(r)$ versus the Euclidean distance for
our coarsest lattice.  As compared to the corresponding plot~\ref{Fig:locality}
using the less-familiar taxi-driver distance, one sees that the data are more
scattered due to violations of rotational symmetry, but are still clearly
contained with a worst-case decay rate.  

\begin{table}[ht]
  \begin{center}
    \begin{tabular}{ll}
      $a$   &  $r_{\rm ov}$ \\
      \hline
      $0.20\,{\rm fm}$ & $1.05(1)$ \\
      $0.17\,{\rm fm}$ & $0.98(1)$ \\
      $0.13\,{\rm fm}$ & $0.9(1)$ \\
      \hline
    \end{tabular}
    \caption{\label{Table:Euclidean} The range (Euclidean metric) at three
             values of lattice spacing. It is less than about 1 lattice unit
             with lattice spacing as coarse as $0.20\,{\rm fm}$. }

  \end{center}
\end{table}

\pagebreak

Again we fit the tail of the function with a decaying exponential to extract
the range.  We tabulate the results in Table~\ref{Table:Euclidean}.  Note that
although the two ranges (taxi-driver and Euclidean) differ by a factor of two
they are quite compatible heuristically; on a $L^4$ hypercube, the maximum
taxi-driver distance is $4L$, and the maximum Euclidean distance is
$\sqrt{4L^2}=2L$.

So even at lattices as coarse as $a=0.2\,{\rm fm}$, the range is about 1
lattice unit (measured using Euclidean distance, or 2 units using taxi-driver
distance).  No unphysical degrees of freedom are induced at distances longer
than the lattice cutoff.

\section{Scaling}

At Lattice 2004, Davies {\it et al.}~\cite{Dav04} collected world data to
demonstrate that different quenched quark formulations could have a consistent
continuum limit.

\begin{figure}[ht]
  \vspace{0cm}
  \begin{center}\hspace*{2cm}
  \includegraphics[clip,angle=0,width=1.0\hsize]{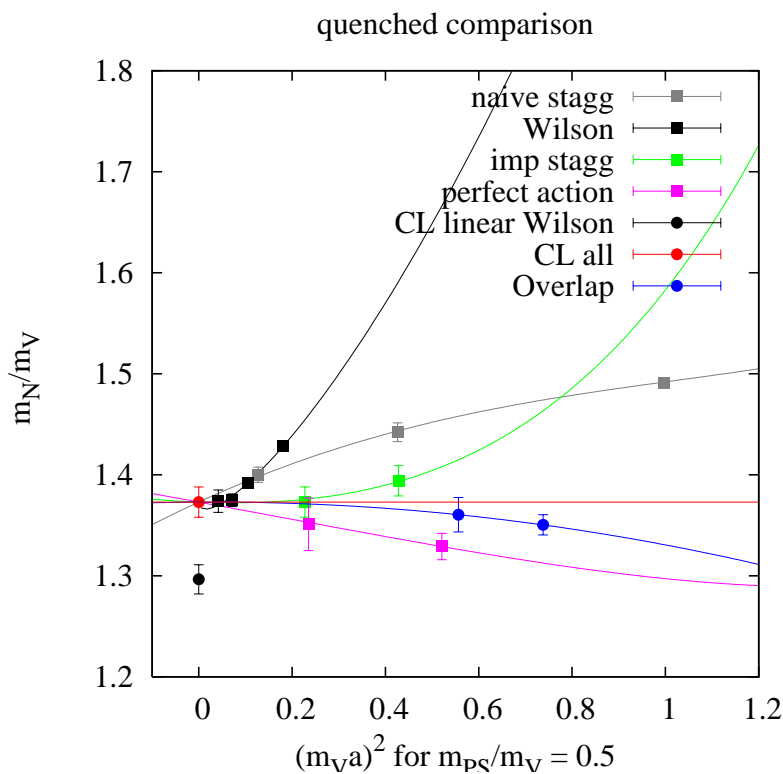}
  \vspace{-0.5cm}
  \caption{\label{Fig:Aoki} ``Aoki'' plot for various quenched data, as
           obtained from~\cite{Dav04}.  Our data is labeled ``overlap''.}
  \vspace{0cm}
  \end{center}
\end{figure}

The conclusion, as illustrated in Fig.~\ref{Fig:Aoki}, is that they could.  But
we emphasize that the constrained fit demanded that there exist a global
continuum limit (by design).  Furthermore, the global continuum limit differed
substantially from the continuum limit obtained solely from the Wilson
formulation where the discretization errors are largest.  The lesson learned is
that with large discretization errors, it is quite possible to be misled when
extrapolating to the continuum limit, even with high statistics and many
lattice spacings.  It is important to seek a formulation with very small
discretization errors to be able to trust the continuum extrapolation.

Here we add our data ``overlap'' to the global quenched spectrum data and find
that of all the formulations, its discretization errors are smallest, allowing
for viable computation at surprisingly coarse lattice spacing.

As another example of the efficacy of the overlap formulation, we made a
non-perturbative computation~\cite{Zha05} of the renormalization constants of
composite operators on the $16^{3}\times 28$ lattice using the regularization
independent scheme.

We found that the relations $Z_A=Z_V$ and $Z_S=Z_P$ agree well (within 1\%)
above $m=1.6\,{\rm GeV}$.  The $m_{{\Lambda}_{\rm QCD}}a^2$ and $(ma)^2$
corrections of the renormalization are small; the mass dependence is less than
about 3\% up to $ma=0.6$.  This is much superior to the competition.

\section{Conclusions}

It is viable to simulate quenched overlap fermions at surprisingly coarse
lattice spacing.  Locality is well under control; the range (characteristic
exponential decay length) is about one lattice unit (of Euclidean distance, or
about two lattice units of Taxi-driver distance) for lattice spacing as coarse
as $0.20\,{\rm fm}$ (such as in~\cite{Che04}), and trends to zero (in physical
units) in the continuum limit.  Scaling is remarkable.  The Aoki plot is
essentially flat up to $0.20\,{\rm fm}$.  The overlap fermion outperforms all
other formulations; discretization errors are smallest for overlap.
Non-perturbative renormalization of operators show little mass
dependence~\cite{Zha05}; e.g.\ less than about 3\% up to $ma=0.6$ for the
renormalization constants

\end{document}